\documentclass{aa}
\usepackage{graphicx}

\begin{document}

\title{Neutrinos from active galactic nuclei as a diagnostic tool} 
\author{C. Schuster, M. Pohl \and R. Schlickeiser}
\offprints{C. Schuster}
\institute{Institut f\"ur Theoretische Physik, Lehrstuhl IV: 
Weltraum- und Astrophysik, Ruhr-Universit\"at Bochum, Germany    
              \\
              \email{cs@tp4.ruhr-uni-bochum.de}         
           }

\date{Received October 29, 2001; accepted November 27, 2001}

\abstract{
Active galactic nuclei (AGN) are known  as sources of  high energy
$\gamma$-rays. The emission probably results from  non-thermal 
radiation of relativistic jets belonging to the AGN. 
Earlier investigations  of these processes  have suggested that 
neutrinos are among the radiation products of the jets  
and may be used to discriminate between hadrons and leptons as 
primary particles for the production of the  high energy emission.  
Our calculation of the high energy neutrino emission from the
jets of AGN  is based on a recently published  model for 
$\gamma$-ray production by a collimated, relativistic blast wave, 
in which the spectral evolution of energetic particles is 
determined by the interplay between the particle injection by 
sweep-up of the interstellar medium, the energy losses through 
radiation, and diffusive escape. It is important to note that 
the swept-up interstellar particles retain their relative velocities 
with respect to the jet plasma, but get isotropised in the jet rest
frame by self-excited $\rm Alfv\acute{e}nic$  turbulence.
The bulk of the neutrino emission is expected in the energy range 
between 100 GeV 
and a few  TeV.
It is shown that the neutrino flux is correlated with the flux 
of TeV $\gamma$-rays. This allows to distinctly search for neutrino 
emissions from the jets of AGN by using the TeV $\gamma$-ray light 
curves to drastically reduce the temporal and spatial parameter 
space. Given the observed TeV photon fluxes from nearby BL Lacs the
neutrino signal from AGN may be detectable with future neutrino 
observatories as least as sensitive as IceCube.  
}

\maketitle

\section{Introduction}
\renewcommand{\dbltopfraction}{0.6}
To date, more than 60 blazar-type active galactic nuclei (AGN)
have been detected as emitters of MeV-GeV $\gamma$-rays by EGRET 
(von Montigny et al. \cite{cvm95}, Mukherjee et al. \cite{muk97}), 
and another few BL Lacertae objects have been observed as sources
of TeV $\gamma$-rays by ground-based \v Cerenkov telescopes (Catanese \&
Weekes \cite{cw99}). In these sources the bulk of luminosity is 
often emitted in the form of $\gamma$-rays. The $\gamma$-ray emission 
of the typical
blazar is highly variable on all timescales down to the observational
limits of days at GeV energies and hours at TeV energies
(e.g. Mattox et al. \cite{mat97}, Gaidos et al. \cite{gai96}).
Emission constraints
such as the compactness limit and the Elliot-Shapiro relation (Elliot \&
Shapiro \cite{es74}) are violated in some $\gamma$-ray blazars, even 
when these limits are calculated in the Klein-Nishina limit 
(Pohl et al. \cite{poh95}, Dermer \& Gehrels \cite{dg95}), which implies 
relativistic bulk motion within the sources. This conclusion is further 
supported by observations of apparent superluminal motion in many 
$\gamma$-ray blazars (e.g. Pohl et al. \cite{poh95}, Barthel et al. 
\cite{bar95}, Piner \& Kingham \cite{pk97a,pk97b}). 

Most published models for the $\gamma$-ray emission of blazars are based
on inverse Compton scattering of soft target photons by highly relativistic
electrons in the jets of these sources. The target photons may come directly
from an accretion disk (Dermer et al. \cite{dsm92},
Dermer \& Schlickeiser \cite{ds93}), or may be rescattered accretions disk
emission (Sikora et al. \cite{sbr94}), or may be produced in the
jet itself via synchrotron radiation (e.g. Bloom \& Marscher \cite{bm93} and 
references therein). The criticism of these models focuses on the fact that an
electron acceleration process would have to be
very fast to compete efficiently with the radiative losses at
high electron energies, and it is therefore unlikely that the
electrons are the primary particles.

The radiating electrons can also be secondary particles produced in
inelastic collisions by primary hadrons, in which case neutrinos would
be emitted in parallel to the $\gamma$-rays. Neutrinos can therefore be used to
distinguish between purely leptonic models and models which involve
high energy hadrons as primary particles.

One class of hadronic models is based on proton acceleration to 
ultra high energies around $10^{20}\ $eV, so that photomeson production 
on ambient target photons can provide many secondary electrons and positrons 
(Kazanas \& Ellison \cite{ke86}, Sikora et al.
\cite{sik87}, Mannheim \& Biermann \cite{mb92}). Such systems are usually
optically thick and a pair cascade develops. These models are particularly
attractive because they allow a possible link be made between escaping ultra 
high energy hadrons and the cosmic ray spectrum at $10^{20}\ $eV near earth
(e.g. Waxman \cite{wax95}). Steady-state calculations indicate that
blazars would produce a high flux of neutrinos with 
energies up to $10^{19}\ $eV (Mannheim \cite{man95}, Halzen \& Zas
\cite{hz97}, Protheroe \cite{pro97}). Note, however, that some of these 
results appear to be in conflict with physical upper limits for the
total neutrino intensity produced by ultra high energy hadrons in AGN
(Waxman \& Bahcall \cite{wb99}, Bahcall \& Waxman \cite{bw01}).
It is unfortunate that time-dependent
calculations of proton-induced cascades have never been performed. We can
therefore not comment on possible correlations between the neutrino flux
of a source and its variable GeV-TeV $\gamma$-ray emission, for the latter 
is also influenced by the presumably variable pair-pair optical depth.

Two of us have recently published a new model of particle energization
in jets of AGN, which is motivated by the blast wave models of gamma-ray bursts
(Pohl \& Schlickeiser \cite{ps00}). There the AGN jet is assumed to be
a cloud of dense plasma which moves relativistically through the interstellar 
medium of the AGN host galaxy. It was shown that
swept-up ambient matter is quickly isotropised in the blast wave frame
by a relativistic two-stream instability, which provides relativistic
particles in the jet without invoking any acceleration process. The typical
primary particle would be a proton with a Lorentz factor of the order of 
one hundred, distributed isotropically in a jet which also moves with a
bulk Lorentz factor of the order of 
one hundred. Inelastic proton-proton collisions with the background plasma
would then lead to the production of TeV $\gamma$-rays by $\pi^0$-decay as 
well as synchrotron, inverse Compton, and bremsstrahlung emission produced
by secondary electrons at lower energies. The fully time-dependent
calculations show that the observed variability behavior and $\gamma$-ray 
spectra of blazars can be explained with this model.

In this paper we are concerned with the neutrino emission which is produced
in parallel to the $\gamma$-rays during the decay of the charged pions. 
We shall provide specific predictions for the neutrino flux and spectra in
relation to the TeV $\gamma$-ray emission of AGN. These will allow observers to
a) search for neutrino emission of AGN by using the TeV $\gamma$-ray light 
curves as indicator of activity, and b) perform a decisive test of at least 
one hadronic model of $\gamma$-ray production in AGN.

In Sect. 2 we briefly review the spectral evolution of the energetic protons,
resulting from the interplay between the injection and the losses. 
We discuss some resulting proton spectra and show that the spectral 
evolution of the proton distribution depends on the geometry and 
the physical constitution of the blast wave.
In Sect. 3 we discuss the various decay modes and present the 
the source functions for the decay processes of the charged pions
as well as the subsequent decay of charged muons. 

We also investigate the emission of 
electron anti-neutrinos caused by $\beta$-decay of neutrons produced
in $p+p\rightarrow p+n+X$ reactions. We show that the power radiated 
in thus produced 
$\overline{\nu}_e$ emission is about two orders of magnitude smaller
than the power radiated in the form of electron (anti-)neutrinos arising from
the decay of muons. In Sect. 4 we discuss the spectral evolution of the
neutrino emission, induced by the proton spectra presented
in Sect.2. We show that (muon)-neutrino emission at energies around 
1 TeV is correlated with TeV $\gamma$-emission.

\section{The energy distribution of protons in an AGN jet}

\begin{figure}[h]
  \centering   
  \resizebox{\hsize}{!}
    {\includegraphics{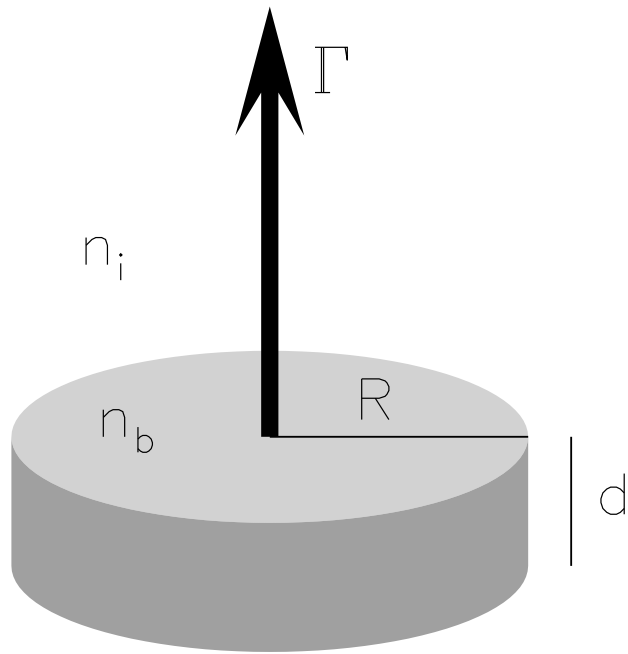}}
  \caption{Sketch of the basic geometry. The thickness of the
channeled blast wave $d$, measured in its rest frame, is much smaller than 
its halfdiameter. The blast wave
moves with a bulk Lorentz factor $\Gamma$ through ambient matter
of density $n_{\rm i}$.}
  \label{geo}
\end{figure}

The AGN model of Pohl \& Schlickeiser (\cite{ps00}) 
describes the jet as a plasma could consisting  of electrons and protons, 
that is assumed to have a cylindrical shape with thickness 
$\rm{d}$, where  $\rm{d}$ is small compared to the radius  
$\rm{R}$ (see also Fig.\ref{geo}). 
The jet is moving with a bulk Lorentz factor $\Gamma$ through 
the ambient medium. Viewed in the rest frame of the jet plasma cloud,
the interstellar medium forms an electron-proton beam of density\footnote{All 
quantities indexed with  ``$\ast$'' \, are in the
laboratory (galaxy) frame.}
$n_{\rm i} =\Gamma\, n_{\rm i}^*$, which is small compared with the plasma
density $n_{\rm b}$ of the jet cloud. Using kinetic theory
Pohl \& Schlickeiser (\cite{ps00}) have shown, that the
incoming particles of the interstellar medium quickly get isotropised 
by self-excited $\rm Alfv\acute{e}nic$ turbulence, but
retain their relative velocities with respect to the jet plasma. 
The jet plasma thus becomes enriched with relativistic particles, of 
which only the protons are of interest for us, for they can produce neutrinos.

\begin{figure*}[t]
\vspace{0.6cm}
\hfill{\includegraphics[bb=200 500 400 700,width=5.5cm]{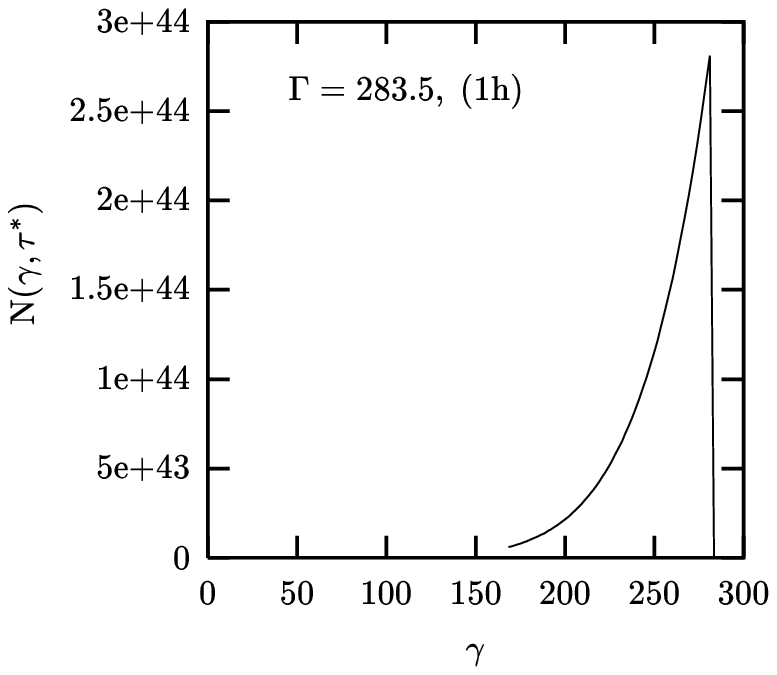}}%
\hfill{\includegraphics[bb=200 500 400 700,width=5.5cm]{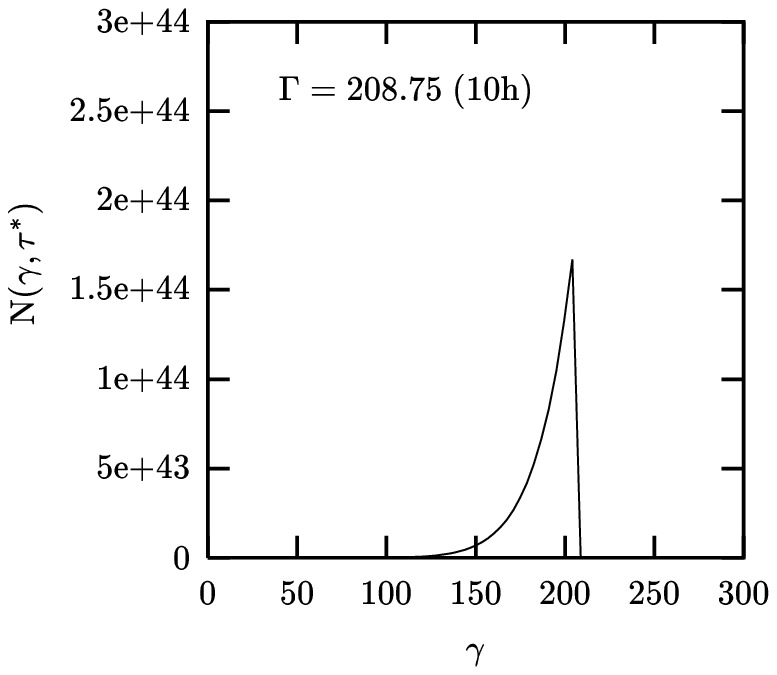}}%
\hfill{\includegraphics[bb=200 500 400 700,width=5.5cm]{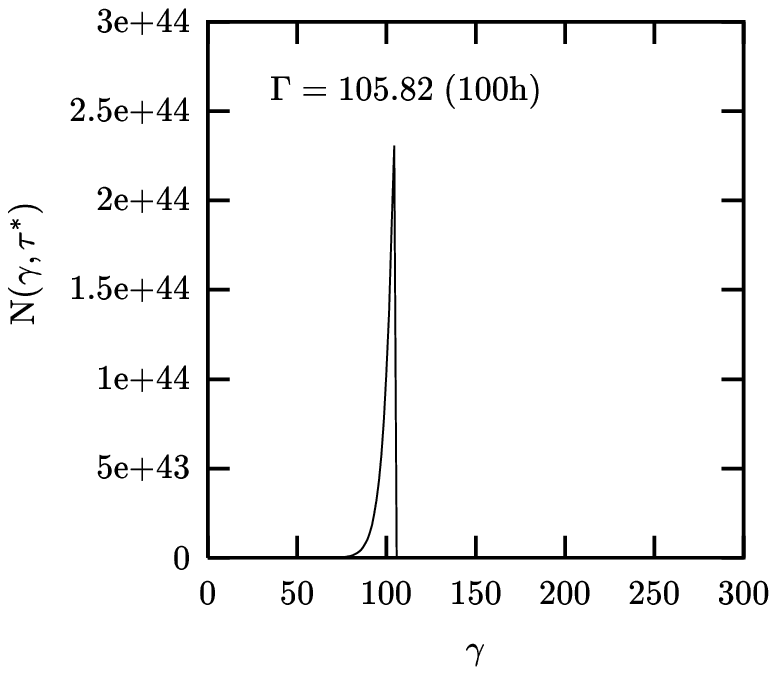}}%
\hspace*{-1.2cm} 
\vspace{-0.3 cm} 
\caption{ A typical spectral evolution for  protons in the blast 
wave, responsible for the neutrino emission. 
In this example  the  following parameters  have been used: 
The radius of the plasma disk is $R = 10^{14}\, \rm{cm}$,\,  
the thickness of the disk amounts to   $d = 3 \cdot  10^{13}\, 
\rm{cm}$\, and the initial Lorentz factor is $\Gamma_0 = 300$.\, 
The constant  densities inside and outside  the jet are 
$n_b = 5 \cdot  10^8 \, \rm{cm^{-3}}$ and   $n_i^\ast = 0.2 \, 
\rm{cm^{-3}}$, respectively.
Note that the mentioned time refers to the observer frame and  
therefore depends on the viewing angle $\theta$, which we choose 
to be $0.1^\circ$ in this example. \vspace{0.5cm}} 
\label{specN1}
\end{figure*}

This has two consequences: 

\noindent a) the system is not stationary.
For reasons of momentum conservation the Lorenz factor $\Gamma$  
decreases at a rate 
\begin{equation}\label{Gammapunkt}
\dot \Gamma = -\frac{\pi  \,R^2 \,n_i^* \,c\, m_p}{M(t)} (\Gamma^2 - 1)^{3/2}.
\end{equation}
Note that the relativistic mass 
\begin{equation}\label{ma1} 
M(t) =  \int_1^\infty \gamma \ m_p N(\gamma,t)\ d\gamma
\end{equation}
is obtained from the particle spectrum in the jet plasma 
\begin{equation}\label{ma2}
N(\gamma,t) = V \, ( n(\gamma,t) +n_b \, \delta(\gamma-1-\epsilon)).
\end{equation}
in which,
in contrast to the nomenclature used in Pohl and Schlickeiser 
(\cite{ps00}), we have have included the background plasma with a formal 
thermal energy $\epsilon = \frac{k\,T}{m_p\,c^2} \ll 1.$

\noindent b) relativistic protons are injected in the jet plasma at a rate
\begin{equation}\label{Einschuss}
S(\gamma,t)= \pi R^2 n_i^* c \sqrt{\Gamma(t)^2-1}\, 
\delta(\gamma-\Gamma(t))
\end{equation}
where $\Gamma(t)$ is determined by Eq. \ref{Gammapunkt}.
The temporal evolution of the proton spectra $N(\gamma,t)$ can be described 
by the continuity equation:
\begin{equation}\label{Konti} 
\frac{\partial N(\gamma,t)}{\partial t}+
\frac{\partial(\dot \gamma N(\gamma,t))}{\partial \gamma}+
\frac{N(\gamma,t)}{T_E} + \frac{N(\gamma,t)}{T_N} 
=S(\gamma,t)
\end{equation}
We consider continuous energy losses from elastic and inelastic 
scattering at a rate of
\begin{equation}\label{strahlung} 
-\dot \gamma = 3 \cdot 10^{-16} n_b \frac{\gamma}
{\sqrt{\gamma^2-1}}
+ 7 \cdot 10^{-16} n_b \frac{(\gamma-1)^2}{\gamma+1}.
\end{equation} 
Additionally, one has to consider catastrophic
particle losses arising from diffusive 
escape at the timescale of 
\begin{equation}
T_E = \frac{3 \,d^2}{\lambda \, \beta \, c} \, 
\simeq 1.67 \, 10^{-19} \, \frac{d^2 \, \Gamma \, n_i^\ast}
{\beta \, \sqrt{n_b}} \; \; \rm{s},
\end{equation}
where $\lambda$ is the scattering length and 
$\beta = \sqrt{1-\gamma^{-2}}$. 

Neutrons produced in $p \rightarrow n$ reactions are likely to 
escape the jet at a timescale of  $T_N \simeq 3\cdot 10^{15}\, 
n_b^{-1} \; \rm{s} $ for $\gamma \gg 1$ (cf. Pohl and Schlickeiser 
\cite{ps00}). 

Neglecting energy losses and additional particle losses, the 
continuity equation (\ref{Konti}) becomes solvable 
and we get a $(\gamma^2-1)^{-3/2}$ spectrum for the 
particle distribution function $N(\gamma,t)$.
However, more realistic cases require the use of numerical methods.

We use Eq. (\ref{Gammapunkt}) to obtain the time $\tau^\ast$ at 
which the blast wave would be observed with a particular 
Lorentzfaktor $\Gamma$. 
\begin{equation}\label{time} 
\tau^\ast = - 
\int_{\Gamma_0}^\Gamma  \,d\Gamma '\,  
\frac {\mathrm{M}(\Gamma ' )\, }{C \,
({\Gamma '}^{2} -  1)^{(3/2)}} \, 
\Gamma ' (1 - \beta \cos \theta^\ast) \, ,
\end{equation} 
where  $C = \pi \,R^{2}\,n_i^\ast\,c\,{m_{p}}$ 
and  $\beta= \sqrt{1 -\Gamma^{-2}}$.
For the purpose of transformation to an observer frame one should  
bear in mind  that the energy scales as well with the factor 
$\Gamma( 1 - \beta \cos \theta^\ast)^{-1}$. 
\begin{figure*}[t]
\vspace{0.6cm}
\hfill{\includegraphics[bb=200 500 400 700,width=5.5cm]{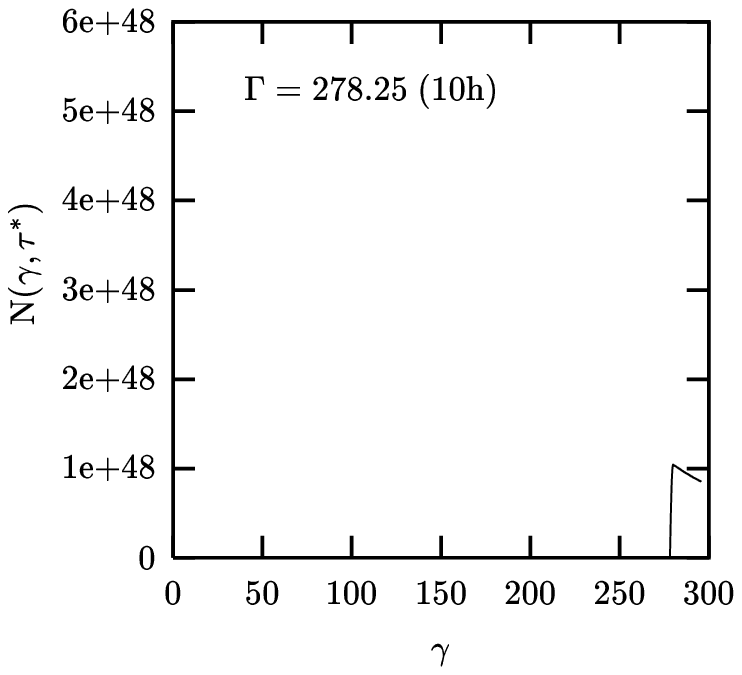}}%
\hfill{\includegraphics[bb=200 500 400 700,width=5.5cm]{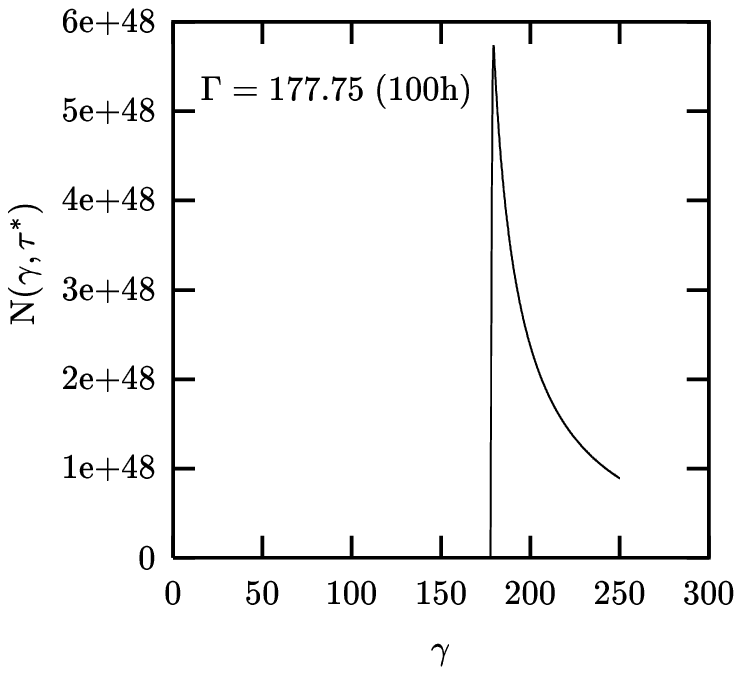}}%
\hfill{\includegraphics[bb=200 500 400 700,width=5.5cm]{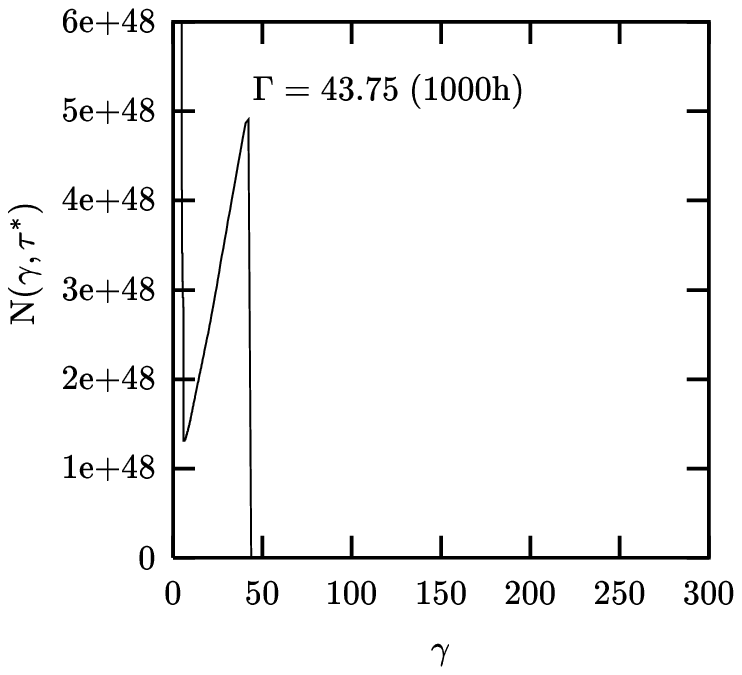}}%
\hspace*{-1.5cm}

\hfill{\includegraphics[bb=200 500 400 700,width=5.5cm]{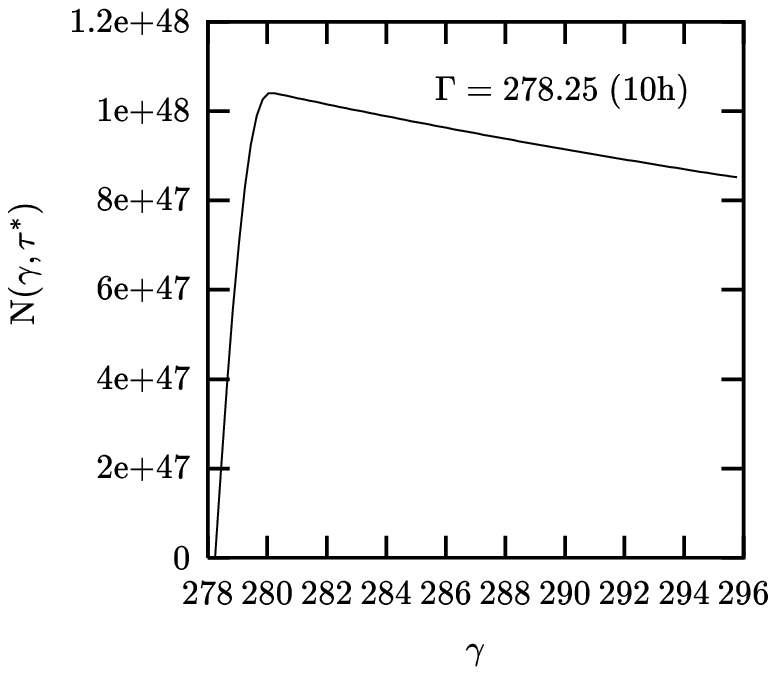}}%
\hfill{\includegraphics[bb=200 500 400 700,width=5.5cm]{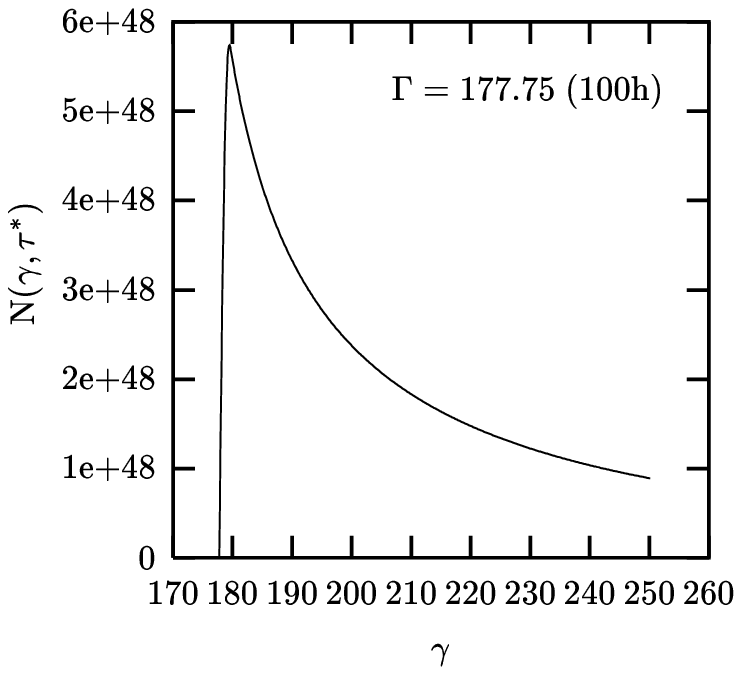}}%
\hfill{\includegraphics[bb=200 500 400 700,width=5.5cm]{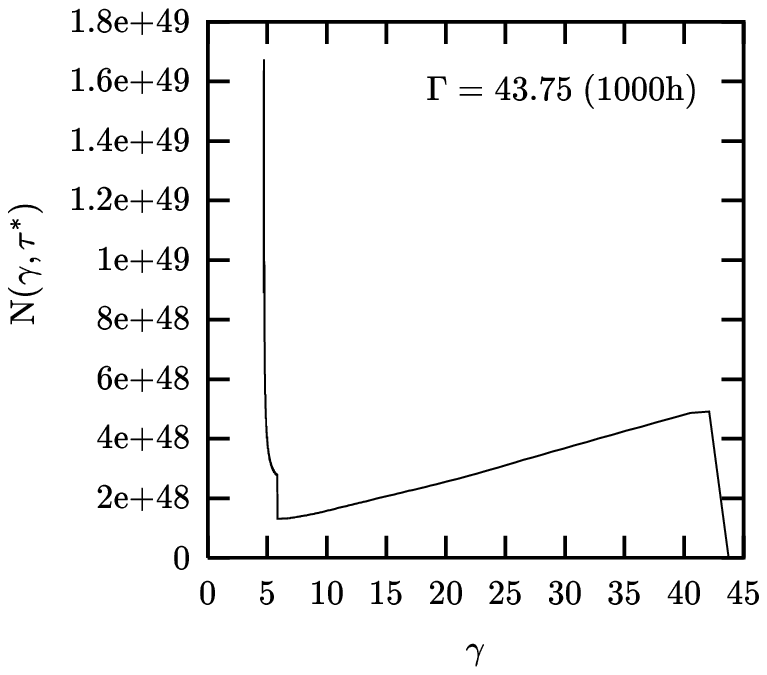}}%
\hspace*{-1.5cm} 
\vspace{-0.3 cm} 
\caption{ The proton spectra calculated with another set of 
parameters. Here we use  $R = 2 \cdot  10^{15}\, \rm{cm}$\,for the
radius of the plasma disk and $d = 10^{14}\, \rm{cm}$\, for  the 
thickness  of the disk. The initial Lorentz factor stays at
$\Gamma_0 = 300$.\, The density in the jet has changed to  
$n_b = 10^8 \, \rm{cm^{-3}}$ and for the density outside the blast 
wave $n_i^\ast = 1.5 \, \rm{cm^{-3}}$ is assumed.
Here the specified time refers to an viewing angle of 
$\theta = 2^\circ$. In the bottom row the same spectra as 
in the top row are shown but with higher resolution.  
\vspace{0.5cm} } 
\label{specN2}
\end{figure*}

In Fig. \ref{specN1} and \ref{specN2} we  calculate typical examples of 
resulting proton distributions $N(\gamma,t)$, describing the number 
density of protons in terms of the Lorentz factor $\gamma$ w.r.t. 
the blast frame.
Starting with an initial Lorentz factor  of $\Gamma_0 = 300$ we show
the distribution of protons for several instances of observed time 
corresponding to the reduced bulk Lorentz factors of the blast wave. 
According to Eq. \ref{Gammapunkt} the Lorentz factor is 
monotonically decreasing. In all cases the sweep-up occurs at the 
present Lorentz factor in accordance with Eq. \ref{Einschuss}.

As mentioned before, the overall behaviour of the particle spectra 
is mainly determined by the competing modes of sweep-up 
and cooling of the particles. 
In Fig. \ref{specN1} the swept-up particles cool down faster than 
the blast wave decelerates, whereas in Fig. \ref{specN2}, 
down to a Lorentz factor of $\Gamma = 85$  the cooling is slow  
compared with the deceleration of the blast wave. 
Afterwards the deceleration rate of the blast wave is overtaken by 
the cooling rate of the particles.
Then the competing mode of particle cooling is the more effective
process. Determined by the initial parameters $d, R, n_i\ast$ and $n_b$
the spectra show a very different behavior. 
The time specified in the pictures refers to the observer frame and 
therefore depends  on the viewing angle $\theta$.

\section{The production of high energetic  neutrinos}

The energetic protons in the jet plasma produce pions in 
inelastic collisions with particles in the background plasma. 
For the calculation of the pion source spectra we use 
the Monte--Carlo model DTUNUC (V2.2) 
(M\"ohring \& Ranft \cite{Moehring}, Ranft et al. \cite{Ranft}, 
Ferrari et al.  \cite{Ferrari}, Engel et al.  \cite{Engel}), 
which is based on a dual parton model  
(Capella et al. \cite{Capella}). 
Here we are concerned with high energetic neutrinos resulting from 
the decays of pions  
$ \pi^\pm \rightarrow \mu^\pm + \,\nu_{\mu}(\overline{\nu}_{\mu})$  
and subsequently
$\mu^\pm \rightarrow e^\pm + \,\overline{\nu}_{\mu}(\nu_{\mu}) 
+ \, \nu_{e}(\overline{\nu}_{e})$. 

\subsection{Pion decay}

\renewcommand{\dbltopfraction}{0.7}
\begin{figure*}[b]
\hspace*{-1cm}
\hfill{\includegraphics[bb=199 490 390 720,width=7cm]{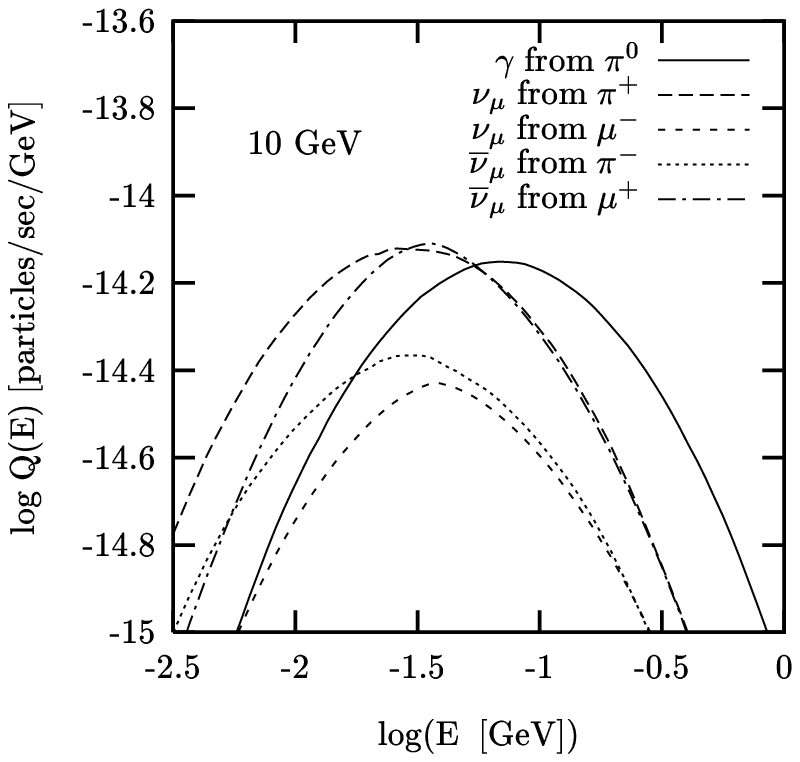}}%
\hfill{\includegraphics[bb=199 490 390 720,width=7cm]{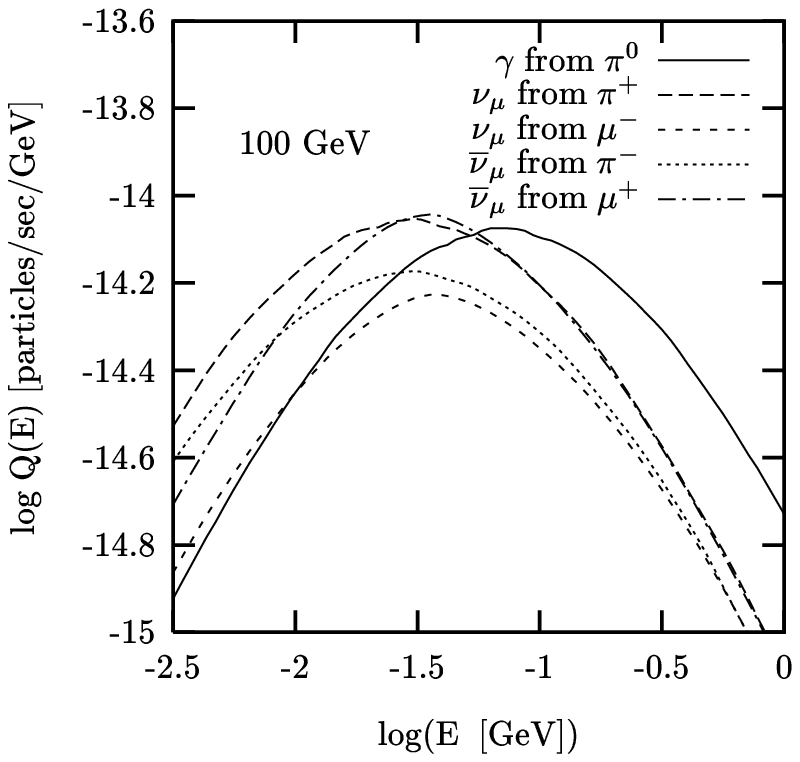}}%
\caption{The production rate of muon neutrinos resulting from the 
various decay modes calculated for  one proton of  kinetic energy 
$10 \;\rm{GeV}$ and $100\; \rm{GeV}$, respectively. 
The emission rate does not change in the energy regime of a proton 
up to  a range of  $  100\; \rm{GeV}$. Additionally we show the 
emission of $\gamma$-rays resulting from the decay 
$ \pi^0 \rightarrow 2 \gamma$ displayed with the solid line. 
The emission from the positively charged pions and muons is slightly
higher than the emission from the negative ones. The respective pion
decay is more effective.} 
\label{spec1}
\end{figure*}
\begin{figure*}[b]
\hspace*{-1cm}
\hfill{\includegraphics[bb=199 490 390 720,width=7cm]{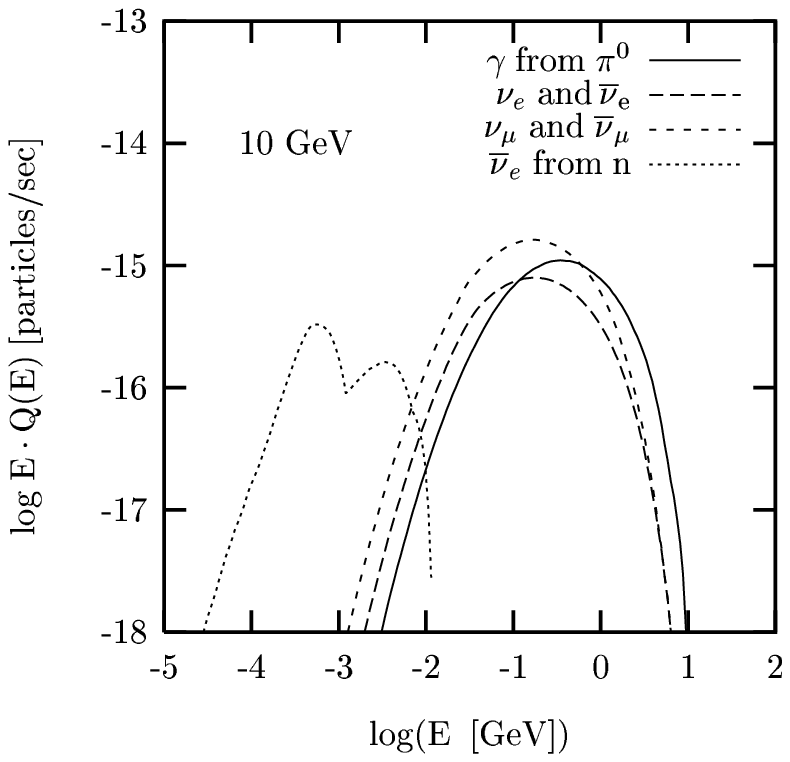}}%
\hfill{\includegraphics[bb=199 490 390 720,width=7cm]{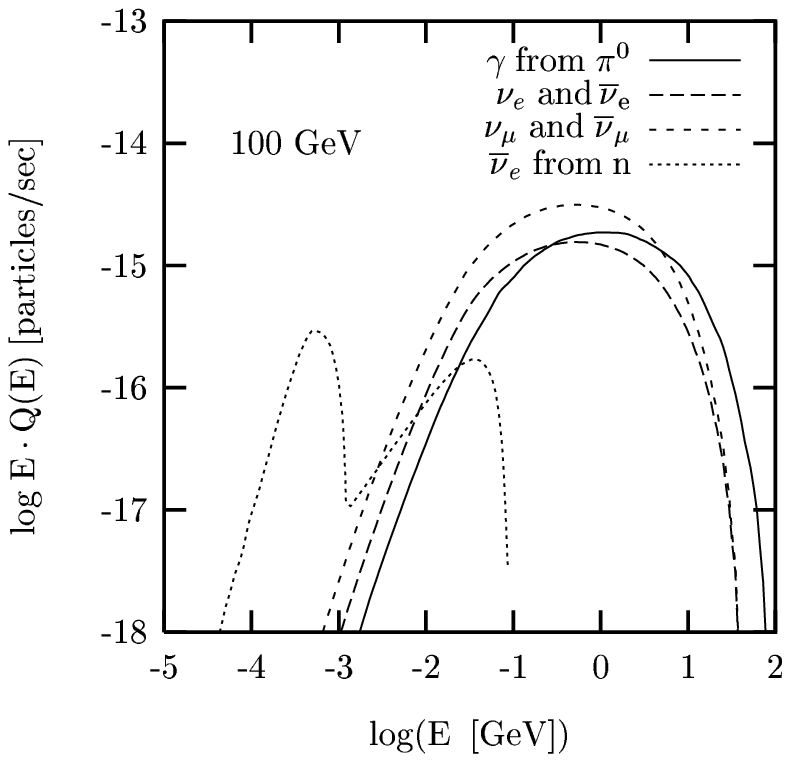}}%
\caption{For a proton with the same kinetic energy as above the 
electron neutrino emission is shown here. The production of 
$\overline{\nu}_e$ and $\nu_e$  resulting from the muon decay is 
displayed by the solid line.  The energy  of anti-electron neutrinos
produced by $\beta$-decay is about 2 orders of magnitude smaller. 
The complete rates of muon neutrinos and $\gamma$-rays are also 
depicted for reference.} 
\label{spec2}
\end{figure*}


The pion decay muon  neutrino source function \,$Q_\nu^\pi(E_\nu,t)$
describing the energy spectrum of the neutrinos produced in the jet
via pion decay  can be written as 
\begin{equation}\label{pq}
Q_\nu^\pi(E_\nu,t) = \int\limits_{\gamma_{\pi \, min}}^\infty   \,  
f(\gamma_\pi) \,  q_{\pi}(\gamma_\pi,t) \; \;  d\gamma_\pi ,
\end{equation}
where $Q_\pi$ is the pion source function and 
\begin{equation}
f( \gamma_\pi )=\frac {1}{2\,{E_{\nu}^0}\,
\sqrt{{\gamma _{\pi }}^{2} - 1}} \, 
\end{equation}
is the normalized neutrino spectrum describing the two-body decay 
in the rest frame of the plasma (Stecker \cite{Stecker}, Marscher 
et al. \cite{Marscher}).
Here $E_\nu^0 \simeq  29.79 \; \rm{MeV}$ 
denotes the energy of the emitted neutrino in the CMS.
The lower limit of integration is given by 
\begin{equation}
\gamma_{\pi \, min} = \frac {E_{\nu}^0}{2\,{E_{\nu }}} + 
\frac {E_{\nu }}{2\,{E_{\nu}^0}}.
\end{equation}
The pion source function 
giving the amount of pions per $\rm{cm^3, \;GeV \;and\;s }$ 
can be calculated as an integral over the spectrum of energetic 
protons and the differential pion 
\clearpage
production cross section which we
derive using the Monte--Carlo code DTUNUC.
\begin{equation}\label{1falt}
q_\pi(\gamma_\pi,t) = \int\limits_1^\infty  
n_b \, N(\gamma_p,t)\, \beta\, c\, 
\frac{d\, \sigma(\gamma_p)}{d\, \gamma_\pi} 
\; d\gamma_p \;\;\; 
\end{equation} 
Here $\beta = v_p/c$ denotes velocity of the protons.
It is convenient to perform the integration over the pion Lorentz 
factor in Eq. \ref{pq} first. Hence
\begin{equation}\label{1afalt}
Q_\nu^\pi(E_\nu,t) = 
\int\limits_1^\infty  
n_b \, \beta\, c\, N(\gamma_p,t)\!\!\!\!\!\! 
\int\limits_{\gamma_{\pi \, min}}^\infty  
\!\!\!\!\!f(\gamma_\pi) \,\frac{d\, \sigma(\gamma_p)}{d\, \gamma_\pi} 
d\gamma_\pi\; d\gamma_p \;\;\; 
\end{equation}
is the differential production rate of particles per 
$\rm{cm^3}$, GeV and s.

\subsection{The neutrino production via decay of charged muons}

In the decay of the pions charged muons emerge producing neutrinos 
in a second decay 
$\mu^\pm \rightarrow e^\pm + \,\overline{\nu}_{\mu}(\nu_{\mu}) 
+ \, \nu_{e}(\overline{\nu}_{e})$.
The muon decay  source functions  for muon and electron neutrinos
$Q_\nu^\mu(E_\nu,t)$ are: 
\begin{equation}\label{muo}
Q_\nu^\mu(E_\nu,t)                            
= \frac {1}{2}\!\!\!{\int\limits _{\gamma _{\mu , \,min}}^{\infty }
\!\!\!\!f(E_\nu) \!\!\int\limits _{\chi_{min}}
^{1}\!\!q_\pi(\gamma _\pi, t) \; 
{\frac {\partial{\gamma _{\pi }} }{\partial{\gamma _{\mu }}
}}\, d\chi d\gamma _\mu }
\end{equation} 
with the neutrino spectrum $f(E_\nu , \gamma_\mu)$ 
in the rest frame of the plasma describing the three-body muon  
decay (Marscher et al. \cite{Marscher}, Zatsepin \& Kuz'min 
\cite{Zatsepin}): 
\parbox{7.5cm}{\begin{eqnarray*}
f(E_\nu)= 16\,{\gamma _{\mu }}^{5}\, (  \, \frac {3}{{
\gamma _{\mu }}^{2}} - \frac {4}{3}(3 + {\beta _{\mu }}^{2})\,\zeta
\,  ) \,\zeta ^{2}\,\frac{1}{m_\mu \;c^2}
\end{eqnarray*}}

\begin{minipage}{0.5cm} for \end{minipage}\quad 
\parbox{6.8cm}{ \begin{eqnarray}
0  \,
\leq \, \zeta \,  \leq \,
\frac {1 - \beta _\mu}{2} 
\end{eqnarray}}

and 
 
\parbox{7.5cm}{\begin{eqnarray*}
f(E_\nu)=\frac {(\frac {5}{3} + \frac {4}{(1 + {\beta _{\mu }})^{3}}
(\frac {8\,\zeta }{3} - 3 (1+{\beta _{\mu }}))\,\zeta ^{2} \, )}{{\beta _{\mu }}\,
\,{\gamma _{\mu } \; m_\mu \;c^2 }}      
\end{eqnarray*}}

\begin{minipage}{0.5cm} for \end{minipage}\quad
\parbox{6.8cm}{\begin{eqnarray}
\frac {1 - \beta _\mu}{2} 
\, \leq \, \zeta \, \leq \,  
\frac {1 + \beta _\mu}{2} \;
\end{eqnarray}}

\noindent for muon neutrinos $ \nu_\mu$ and electron neutrinos $\nu_e$.
The  spectra for the  production of anti-muon neutrinos 
$\overline{\nu}_\mu$ and anti-electron neutrinos 
$\overline{\nu}_e$ are calculated by:

\parbox{7.5cm}{\begin{eqnarray*}  
f(E_\nu)=32\,{\gamma _{\mu }}^{5}\, (  \, \frac {3}{{
\gamma _{\mu }}^{2}} - 2 (3 + {\beta _{\mu }}^{2})\,\zeta
 \,  ) \,\zeta ^{2}\,\frac{1}{{m_\mu \;c^2 }}
\end{eqnarray*}}  

\begin{minipage}{0.5cm} for \end{minipage}\quad 
\parbox{6.8cm}{ \begin{eqnarray}
0  \,
\leq \, \zeta \,  \leq \,
\frac {1 - \beta _\mu}{2} 
\end{eqnarray}}

and 

\parbox{6.8cm}{\begin{eqnarray*}
f(E_\nu)=\frac {(1 + \frac {4}{(1 + {\beta _{\mu }})^{3}}
(4\,\zeta  - 3 (1+{\beta _{\mu }}))\,\zeta ^{2} \, )}
{{2 \beta _{\mu }}
\,{\gamma _{\mu } \; m_\mu \;c^2}} 
\end{eqnarray*}} 

\begin{minipage}{0.5cm} for \end{minipage}\quad
\parbox{6.8cm}{\begin{eqnarray}
\frac {1 - \beta _\mu}{2} 
\, \leq \, \zeta \, \leq \,  
\frac {1 + \beta _\mu}{2} \;
\end{eqnarray}}
where $\zeta = (\frac{E_\nu}{\gamma_\mu  \, m_\mu \, c^2})$ 
and $\beta_\mu = v_\mu/c.$
Here $\gamma_\pi(\gamma_\mu, \chi)$ denotes the Lorentz factor of 
the parent pion: 
\begin{equation}
{\gamma _{\pi }}=\frac {{\varepsilon _{2}}\,{\varepsilon _{1}}
 + \chi \,\sqrt{{\varepsilon _{2}}^{2} - {\varepsilon _{1}}^{2}
 + \chi ^{2}}}{{\varepsilon _{1}}^{2} - \chi ^{2}} ,
\end{equation}
where  ${\varepsilon _{1}}= 3.68 $, \,
${\varepsilon _{2}}= 3.55 \, \gamma_{\mu}$ and $\chi$ 
is the CMS  interaction angle cosine. 

\noindent The lower limits of integration of (\ref{muo}) are:

\begin{equation}
\gamma _{\mu,\; min}= \frac {E _\nu \, }{m_\mu \, c^2} + 
\,  \frac {m_\mu \, c^2}{4\,E _\nu }
\end{equation} 
       
and
 
\[  \chi _{min}   =   \left\{ \begin{array}{c@{\quad:\quad}l}    
 - 1 &  {\varepsilon _{1}}\leq {\varepsilon _{2}} \\
\sqrt{{\varepsilon _{1}}^{2} - {\varepsilon _{2}}^{2}} &
{\varepsilon _{1}} > {\varepsilon _{2}}        
\end{array} \right. \] 
The calculation of the source  function 
$Q_\pi(\gamma _\pi (\gamma _\mu, \chi ), t) $ 
is based on the Monte Carlo code DTUNUC again 
and is calculated analogous to the previous case with 
formula \ref{1falt}.

\begin{figure*}[t]
\centering
{\includegraphics[bb=148 510 370 715,width=5.9cm]{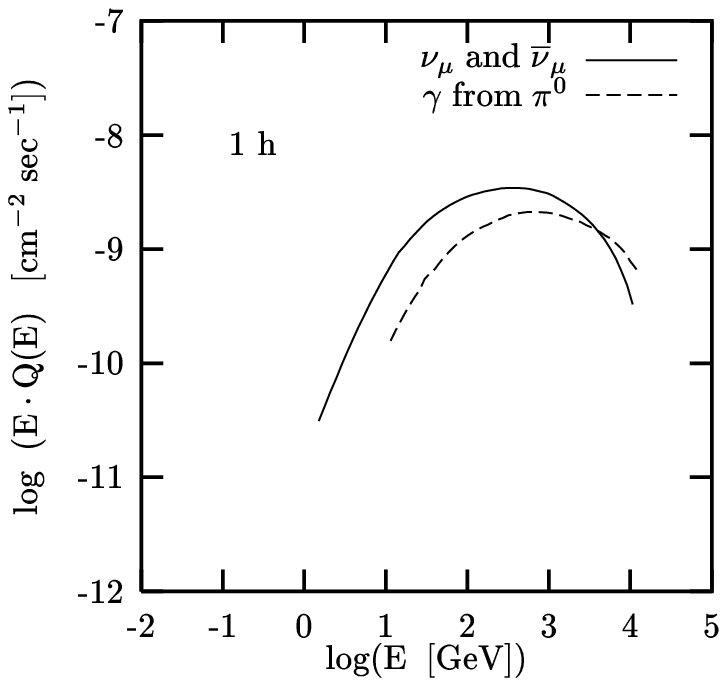}}%
{\includegraphics[bb=143 510 365 715,width=5.9cm]{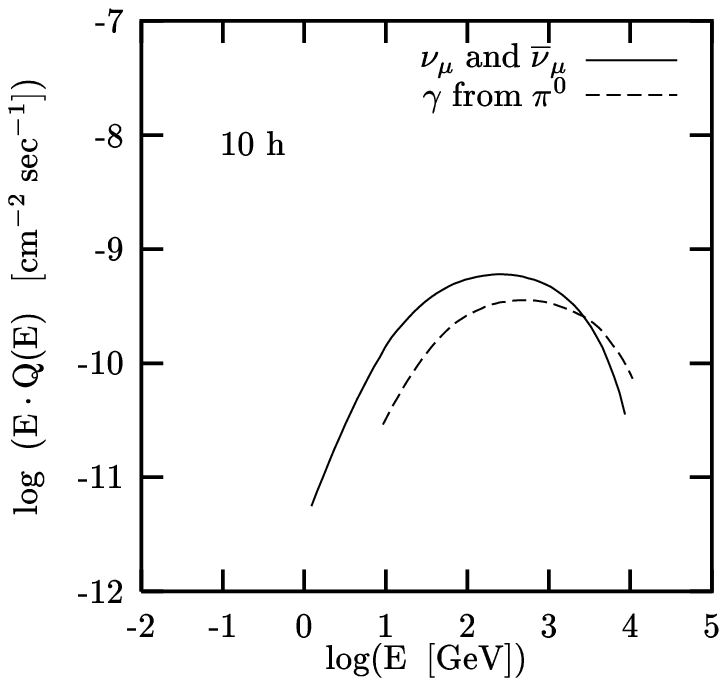}}%
{\includegraphics[bb=138 510 360 715,width=5.9cm]{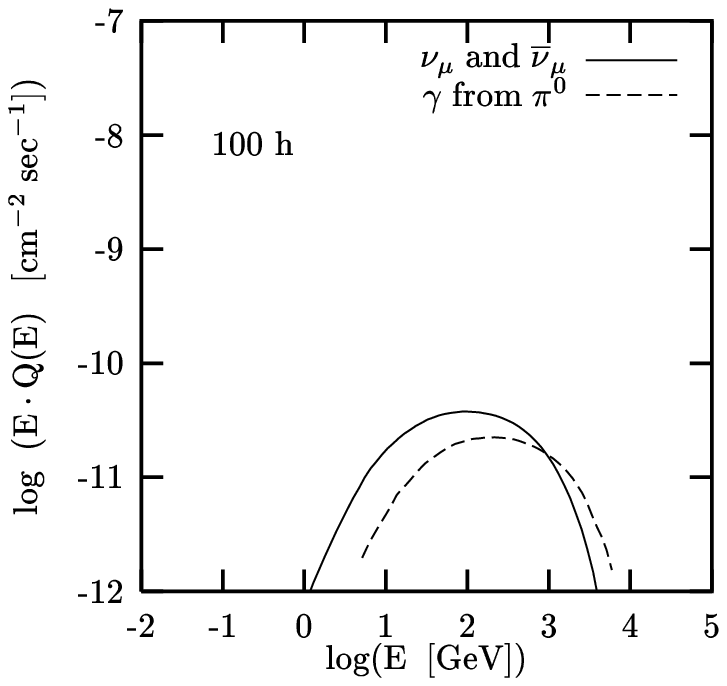}}%
\vspace{0.3 cm} 
\centering
{\includegraphics[bb=148 510 370 715,width=5.9cm]{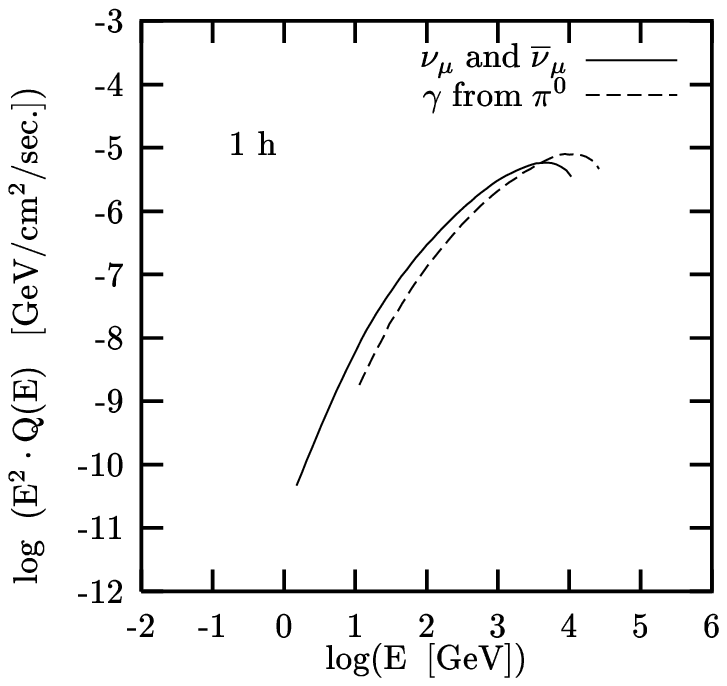}}%
{\includegraphics[bb=143 510 365 715,width=5.9cm]{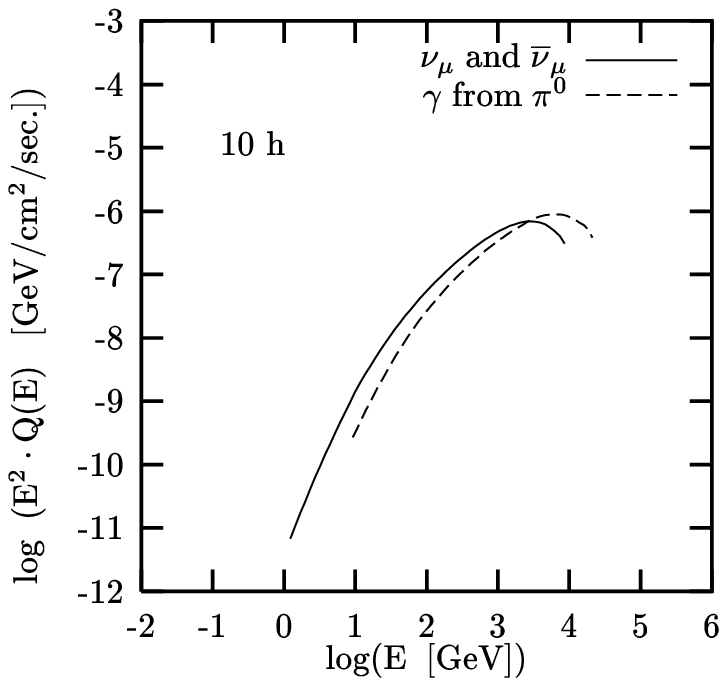}}%
{\includegraphics[bb=138 510 360 715,width=5.9cm]{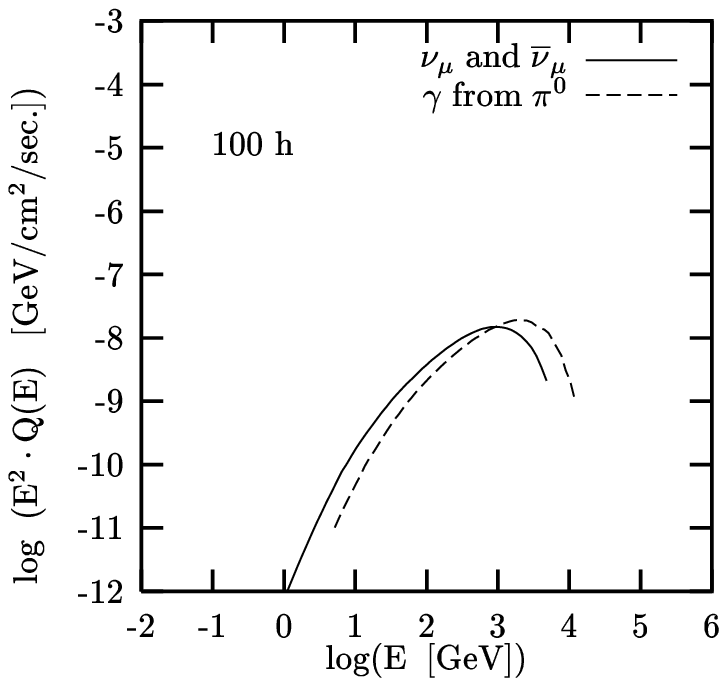}}%
\vspace{0.3 cm} 
\caption{ The evolution of the muon neutrino emission resulting 
from the proton spectra of Fig. \ref{specN1}. The viewing angle 
is $\theta = 0.1^\circ$ and the redshift 
of the AGN is $z=0.5$. All the other parameters are chosen as in 
Fig. \ref{specN1}. In the top row we see the $F_\nu$ spectra and in 
the bottom row the $\nu \,F_\nu$ spectra. The spectral evolution of 
the $\gamma$-ray production spectra is also shown for reference.   
\vspace{0.5cm} } 
\label{spec3}
\end{figure*}

\subsection{Neutron $\beta$-decay } 

The emission of electron anti-neutrinos resulting from the 
$\beta$-decay of secondary neutrons  
$ n \rightarrow  p + e + \overline{\nu}_e $
is described by the electron $\beta$-decay source function 
$Q_{\overline{\nu}_e}(E_\nu,t)$  calculated analogous to 
Jones (\cite{Jones}) 
treating the neutrino as a particle without mass:
\begin{equation}\label{wn}
Q_{\overline{\nu}_e}
= \frac {1}{2} 
\int\limits_{0}^{Q} 
\!\frac{f(E_\nu^0)}{E_\nu^0} \!
\! \int\limits_{\gamma _{ min}}^\infty 
\! \frac{Q_n
(\gamma_n ,t)} 
{\sqrt{{\gamma _{n}}^{2} - 1}}\,d{\gamma _{n}}\;dE_\nu^0 
\end{equation} 
with  the upper limit of integration 
$Q = m_n \; c^2 - m_p\; c^2 $ and  the  lower limit of integration 
$\gamma_{min}$  given as:
\begin{equation}
\gamma_{min} = \frac{1}{2}\;(\frac{E_\nu}{E_\nu^0}+
\frac{E_\nu^0}{E_\nu})
\end{equation}
The  CMS normalized antineutrino spectrum is given as (Leon \cite{Leon}):
\begin{equation}
f(E_\nu^0) =  C \; (Q - E_\nu^0) 
\sqrt{ (Q -  E_\nu^0)^2 - m_e^2 \, c^4} \;{E_\nu^0}^2   
\end{equation}
Again  $E_\nu^0$ denotes the energy of the emitted neutrino in 
the CMS. $C$ is the constant for normalization and is 1.2 $\rm{MeV^{-5}}$. 
For the description 
of the neutron source function  $Q_n(\gamma_n ,t)$ we use a discrete
sum of intensities of monoenergetic neutrons, each one modeled  
as a $\delta$-distribution. So one assumes that the thermal protons 
take part in the collision process as well as the relativistic ones.
The approximation for the source function $Q_n(\gamma_n ,t)$ is:
\begin{equation}                                                   
Q_n = \int\limits_1^\infty  \frac{ N(\gamma_p,t)}{\tau}\,
\left( \delta (\gamma_n - \frac {\gamma}{2} ) + 
\delta (\gamma_n - 1.1) \right)\;d\gamma_p 
\end{equation}
with $\tau =3 \cdot 10^{15} \; n_b^{-1} $.
The second $\delta$-function with the argument 1.1 
accounts for the thermal protons.  
Again we change the order of integration: 
\begin{displaymath}
Q_{\overline{\nu}_e} = \int\limits_1^\infty 
\frac{ N(\gamma_p, t)}{2\;\tau} 
\int\limits_{0}^Q \frac{f(E_\nu^0)}{E_\nu^0}\! 
\int\limits_{\gamma _{min}}^\infty \!\! 
\frac{D(\gamma_n)}{\sqrt{{\gamma _{n}}^{2} - 1}}\; 
d{\gamma _{n}}\;dE_\nu^0\; d\gamma_p  
\end{displaymath}
\begin{equation}  
\hphantom{ Q_{\overline{\nu}_e}(E_\nu,t)=}
\end{equation}
with $D(\gamma_n) =\delta 
(\gamma_n - \frac{\gamma}{2}) + \delta (\gamma_n - 1.1)$. 
To obtain the results in the observer frame one has to 
consider the transformation:  
\begin{equation}\label{transform}
Q^\ast(E_\nu^\ast, \tau^\ast)  = \frac{Q(E_\nu(E_\nu^\ast), 
\tau^\ast)}{ \Gamma^2 \,( 1 - \beta \cos \theta^\ast)^2},
\end{equation}
where all quantities indexed by ``$^\ast$''\, are calculated
in the observer system. 

\begin{figure*}[t]
\vspace{0.1cm}
\hfill{\includegraphics[bb=200 500 400 700,width=5cm]{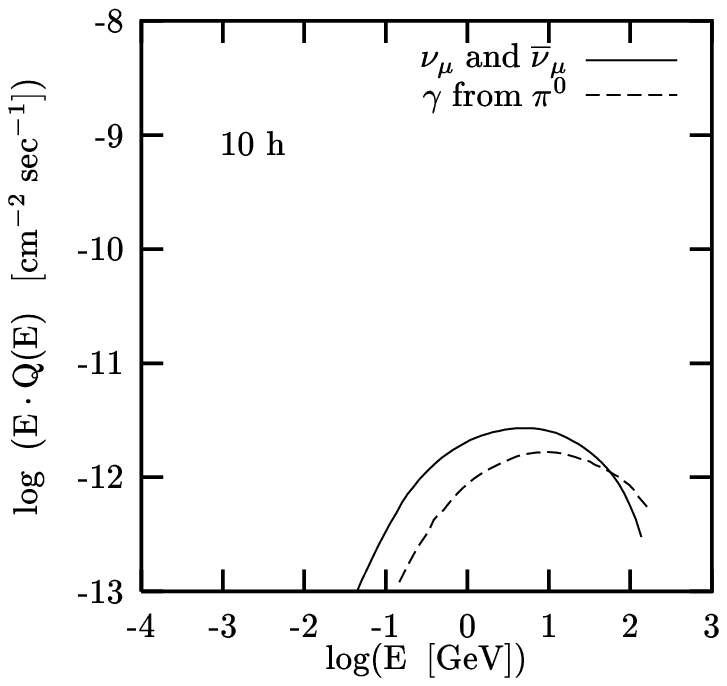}}%
\hfill{\includegraphics[bb=200 500 400 700,width=5cm]{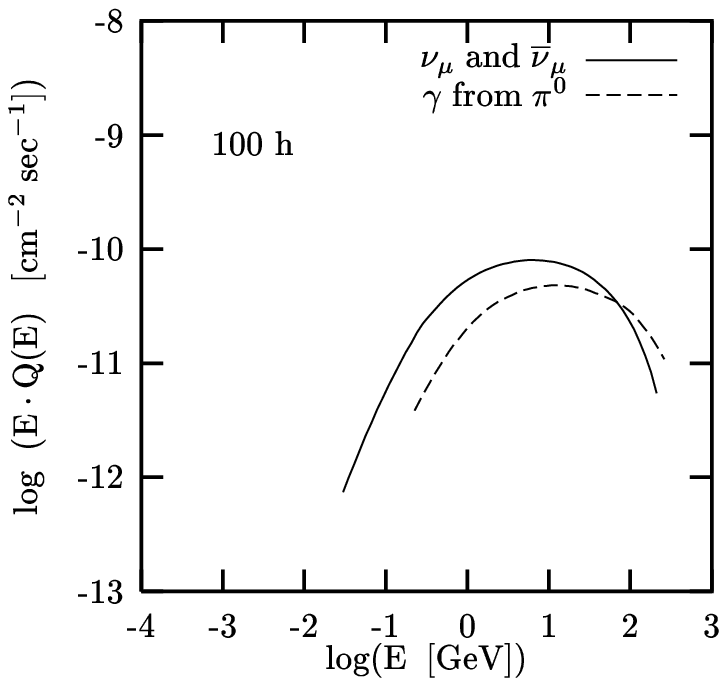}}%
\hfill{\includegraphics[bb=200 500 400 700,width=5cm]{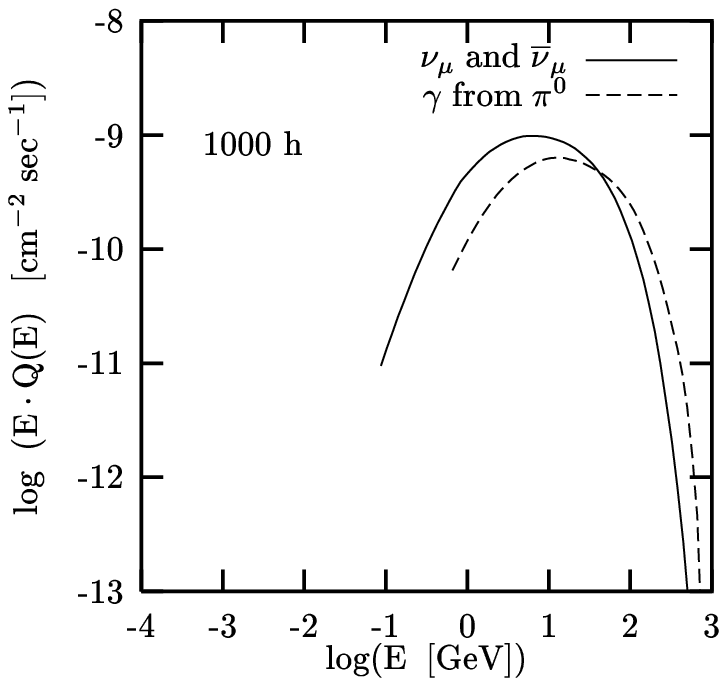}}%
\hspace*{-1.5cm} 

\hfill{\includegraphics[bb=200 500 400 700,width=5cm]{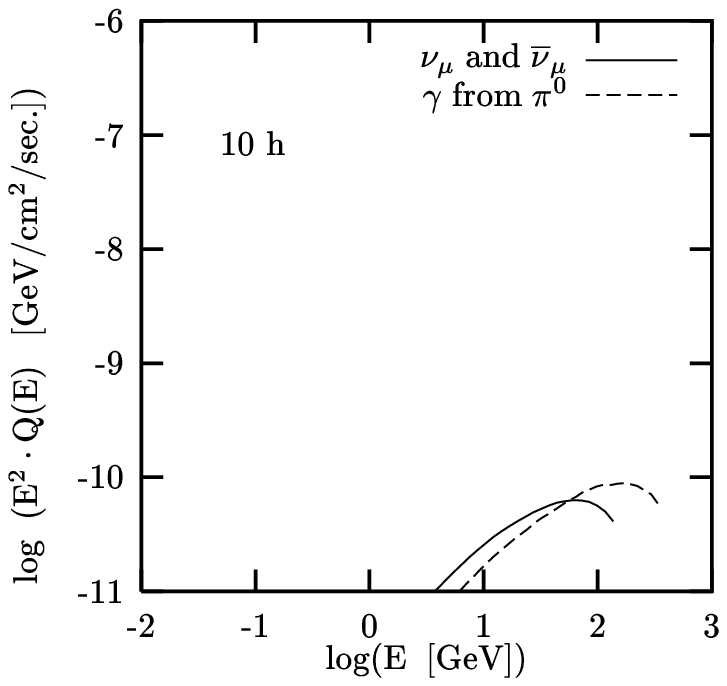}}%
\hfill{\includegraphics[bb=200 500 400 700,width=5cm]{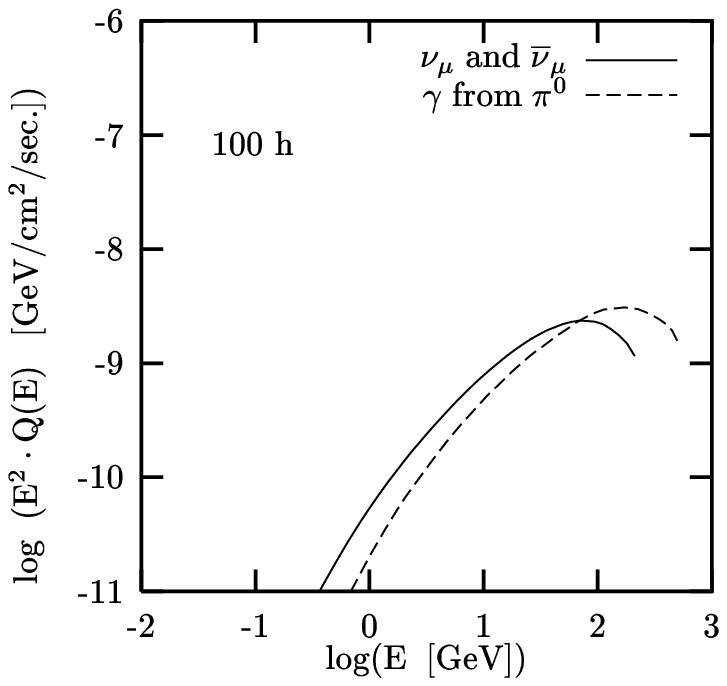}}%
\hfill{\includegraphics[bb=200 500 400 700,width=5cm]{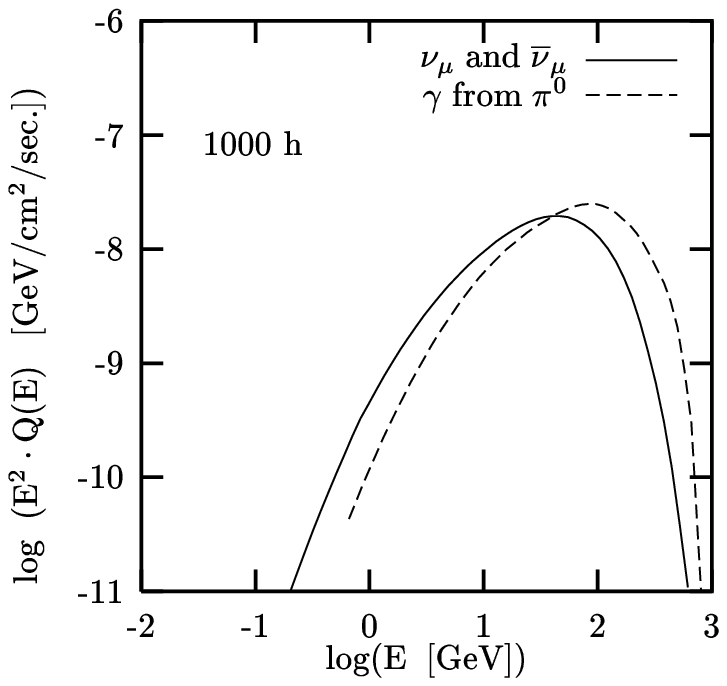}}%
\hspace*{-1.5cm}
\vspace{-0.3 cm} 
\caption{The second set of parameters used in Fig. \ref{specN2} 
results in these spectral evolution. In this example we use the viewing angle  
$\theta = 2^\circ$. The strong dependence of the observed emission
from the  angle $\theta$ is due to the high Lorentz factors in our model.  
This is responsible for the strong rise after 1000 h. 
The set of parameters is the same as in Fig. \ref{specN2} and 
the redshift is again $z=0.5$.  
\vspace{0.5cm} } 
\label{spec4}
\end{figure*}

\section{The evolution of the  neutrino emission spectra  
in comparison with $\gamma$-ray production}

In this paper we focus on the high energy muon neutrino 
emission resulting from the decay mode  of pions  
$ \pi^\pm \rightarrow \mu^\pm + \,\nu_{\mu}(\overline{\nu}_{\mu})$  
and subsequently
$\mu^\pm \rightarrow e^\pm + \,\overline{\nu}_{\mu}(\nu_{\mu}) 
+ \, \nu_{e}(\overline{\nu}_{e})$.  

In Fig. \ref{spec3} and  Fig. \ref{spec4} we show two examples
of spectral evolution of total muon neutrino emission calculated 
with the model proton spectra seen in Fig.\ref{specN2} and in 
Fig.\ref{specN2}. We assume a constant density of the background 
plasma. The production rate of $\gamma$-ray resulting 
from the decay $\pi^0 \rightarrow 2 \gamma$ is depicted as well
for reference. The bulk of muon neutrino emission lies between 
100 GeV and 1 TeV  and follows strictly the evolution of 
$\gamma$-ray production. 
All spectra are calculated using the Monte-Carlo code DTUNUC 
described earlier in this paper.
Emitted $\gamma$-ray photons may be absorbed by collisions with
infrared background photons in the intergalactic medium, but 
neutrino emission is unaffected by any interaction. 
So for known $\gamma$-ray production the 
neutrino emission  may even exceed the calculated rates.

Despite large production rates the detection of high energy 
neutrino emission requires large water or ice telescopes, based 
on detection of \v{C}erenkov light (for example the AMANDA 
experiment at the south pole). In this section we want to 
estimate whether or not neutrino emission from blazars will 
be observable, given their TeV $\gamma$-ray fluxes.

Suppose a TeV source is observed with a $\gamma$-ray flux of
\begin{equation}
N_\gamma \simeq 10^{-10}\ {\rm ph.\,cm^{-2}\,sec^{-1}}\qquad
 {\rm  at\ a\ few\ TeV}
\label{gafluss}
\end{equation}
As we have seen, this implies a similar flux of muon neutrinos at about a
TeV. Following the treatment by Gaisser \& Stanev (\cite{gai84}), we
can calculate the differential flux of muons produced in deep-inelastic 
interactions of the incoming neutrinos with the ice material as
\begin{equation}
{{dN_\mu}\over {dE_\mu}} \simeq {{N_{\rm Av}\,\sigma}
\over \alpha}\,E_\nu\,N_\nu
\approx 10^{-19}\ {\rm GeV^{-1}\,cm^{-2}\,sec^{-1}}
\label{mufluss}
\end{equation}
when the deep-inelastic cross section $\sigma$, the muon energy loss rate
$\alpha$, the Avogadro number $N_{\rm Av}$, and the neutrino flux according
to the $\gamma$-ray flux (\ref{gafluss}) are inserted. 
The muon spectrum is flat for
$E_\mu \ll E_\nu$ and the mean energy $\overline{E_\mu}\simeq 1/3\,
\overline{E_\nu}\simeq 0.3\ $TeV.
Then the total muon flux is approximately
\begin{equation}
N_\mu \simeq \overline{E_\mu}\,{{dN_\mu}\over {dE_\mu}}\simeq 
3\cdot 10^{-17} \, {\rm cm^{-2}\,sec^{-1}}
\label{tmufluss}
\end{equation}
The number of neutrino-caused event now depends on the effective area of the
detector $A_{\rm eff}$, and the observing time $t_{\rm obs}$. The effective
area of the AMANDA detector, which is operational to date, is about 
$A_{\rm eff_A}\simeq 10^4\ {\rm m^2}$, whereas the projected IceCube 
experiment may have $A_{\rm eff_I}\simeq 10^6\ {\rm m^2}$.

The number of events from a TeV $\gamma$-ray source with photon flux 
(\ref{gafluss}) in the two detectors is then
\begin{equation}
N_{\rm A} \simeq 0.01 \ \left({t_{\rm obs}\over {\rm Month}}\right)
\qquad\qquad
N_{\rm I} \simeq 1 \ \left({t_{\rm obs}\over {\rm Month}}\right)
\label{event}
\end{equation}
Equation (\ref{event}) makes clear that the observation of neutrinos from AGN 
is, if at all, only possible with future neutrinos observatories like IceCube. 

Is it possible to identify an AGN neutrino signal in the data of IceCube?
The actual detection of some neutrino events in the detector is not 
sufficient for them being identified as coming from an AGN. We need to observe
a local excess of events on top of the nearly isotropical distribution of
events caused by atmospheric neutrinos. The intensity of atmospheric
neutrinos at 1 TeV is about (Volkova, Zatsepin and Kuzmichev \cite{vzk80})
\begin{equation}
I_{\nu_{\rm atm}} \simeq 10^{-7}\ {\rm cm^{-2}\,sec^{-1}\,sr^{-1}}\qquad
 \rm{at} \ 1\ {\rm TeV}
\label{I_nu_atm}
\end{equation}
It varies by a factor of five between vertically and
horizontally incident particles, but can be taken as locally flat.
At muon energies around 300 GeV the muon follows the flight direction to within
slightly more than one degree. No matter how well the detector can reconstruct
the muon flight direction, the uncertainty in the reconstructed neutrino 
flight direction will in any case be worse than one degree. For our purpose
it may suffice to assume that the neutrino incidence direction can be 
reconstructed to within two degrees, which leaves some room for 
uncertainties in the muon event reconstruction. We can then define a 
resolution element, $\Omega\simeq 3.8 \cdot 10^{-3}\ {\rm sr}$, and the 
number of events of atmospheric neutrinos per resolution
element $N_{\rm atm}$ for Icecube as  
\begin{equation}
N_{ atm}= I_{\nu_{\rm atm}}\,\Omega \; A_{\rm eff_I} \; t 
\simeq  4 \; \left({t_{\rm obs}\over {\rm Month}}\right) 
\quad \rm{at}\ 1\ {\rm TeV}
\label{N_atm}
\end{equation}
which is four times the expected event number of neutrinos 
from an AGN with $\gamma$-ray flux (\ref{gafluss}).

The expected number $N_{\rm AGN}$ of atmospheric neutrino events per 
resolution element would be well defined, for we can use thousands of 
resolution elements to determine it. Given an observed number of events, 
$N_{\rm obs}$, we would deduce the number of events due to AGN neutrinos as 
\begin{equation}
N_{\rm AGN}= N_{\rm obs}-N_{\rm atm}\simeq 
N_{\rm obs}- 4 \; \left({t_{\rm obs}\over {\rm Month}}\right)
\label{dreiunddreissig}
\end{equation}

In the null hypothesis $N_{obs}$ follows a Poissonian distribution.
\begin{equation}\label{poisson} 
P(N_{obs},N_{atm}) = \frac{N_{atm}^{N_{obs}}}{N_{obs}!} \; exp(-N_{atm})
\end{equation}

\begin{figure}[t]
\centering
{\includegraphics[bb=80 135 555 630,width=8.4cm]{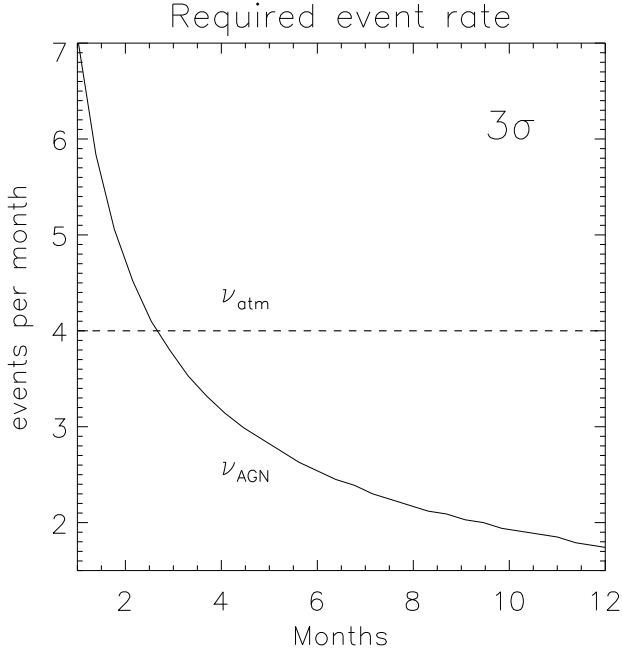}}%
\vspace{0.3 cm} 
\caption{The required rate of AGN neutrino events in IceCube for a
$3\sigma$ excess per resolution element as a function of the observing time.
The neutrino rate expected for a TeV $\gamma$-ray source with a $\gamma$-ray 
flux as in Eq.~\ref{gafluss} would be one per month. Therefore an observing 
time of nearly one year would be needed to produce a $3\sigma$ signal.
\vspace{0.5cm} } 
\label{excess}
\end{figure}

Equation (\ref{dreiunddreissig}) allows us to determine the event rate 
of AGN neutrinos required to provide a signal of specified significance 
by appropriate summation of the Poissonian probability distribution 
(\ref{poisson}). 
As an example we give the result for a $3\sigma$ excess in Fig~\ref{excess}.
Please note that the significance applies for an excess in an 
predetermined resolution element.

Apparently much more than 10 events are necessary to detect signals
from AGN.  By approximating the width of the Poissonian distribution by 
the square root of its mean we can therefore derive a simple formula 
to calculate the signal to noise ratio as a function of the exposure 
time $t$, effective area $ A$ and the
$\gamma$ source flux $N_\gamma$ from AGN:
\begin{equation}
n \simeq \frac{1}{2} \; \left (\frac{N_\gamma}
{10^{-10} \; {\rm cm^{-2} s^{-1} }}\right ) \;\;
{\rm \left (\frac{A_{eff}}{10^6 m^2}\right )^{1/2} 
\left ( \frac{t_{obs}}{Month} \right )^{1/2}}
\label{n_agn}
\end{equation}
  
Mkn501 has been observed to emit a  $\gamma$-ray flux of 
$0.5 \cdot 10^{-10} {\rm cm^{-2} \; s^{-1}}$ at a few TeV
for about half a year (Quinn et al. \cite{qin}).
 Mkn421 has been similarily active in the first half of 2001,
as discussed at the ICRC 2001.
So this indicates the flare states of many AGN would have to be 
co-added to expect a meaningful neutrino signal in the IceCube
detector.
Nevertheless, because forthcoming TeV Cherenkov telescopes 
will provide much better sampled  light curves of many more AGN,
the detection of a neutrino signal from AGN is still possible.   

\section{Discussion}

We have calculated the neutrino emission resulting from jets of
AGN, under the assumption of a channeled blast wave model.
Therefore we have investigated the decay modes of pion and 
subsequent muon decay. Neutrino emission resulting from the 
decay of secondary neutrons has been described as well. We have shown
that neutrinos resulting from $\beta$-decay of neutrons
possess an energy about two orders of magnitude lower than
that of the other decay channels we considered.  
The bulk of the neutrino emission is expected to be in the energy 
range between 100 GeV and 1 TeV.

The emission rate resulting from a proton pair in the 
energy regime between $1$ GeV and $100$ GeV does not vary much.
Therefore the time dependence of the emission spectra is
completely determined by the changing spectra of incoming protons.
We additionally discuss some spectra resulting from the channeled 
blast wave model of Pohl and Schlickeiser (\cite{ps00}) and observe 
that their shape is determined by the ratio between the cooling rate
of the particles in the blast wave and the deceleration rate of the 
blast wave itself.   

Neutrino emission should  be correlated with the  emission of 
$\gamma$-rays. It allows to distinctly look for  neutrino emission 
from the jets of AGN by using the TeV $\gamma$-ray light curves to 
drastically reduce the temporal and spatial parameter space in the 
search for neutrino outbursts. Given the observed TeV photon fluxes 
from nearby BL Lacs the neutrino flux can be detectable with future 
neutrino observatories.
 
\begin{acknowledgements}
Partial support by the Bundesministerium f\"ur Bildung und Forschung through 
the DESY, grant {\it 05 CH1PCA 6}, is gratefully acknowledged.
\end{acknowledgements}

\end{document}